\documentclass[%
 preprint,
groupedaddress,
 amsmath,amssymb,
 aps,
pre,
]{revtex4-1}
\usepackage{enumerate}
\usepackage{graphicx}
\usepackage{dcolumn}
\usepackage{bm}
\usepackage{graphicx}
\usepackage{bbold} 
\usepackage{graphicx}
\usepackage{dcolumn}
\usepackage[left=3cm,top=2.5cm,right=3cm,bottom=2.5cm]{geometry}
\usepackage{array}
\usepackage{epsfig}
\usepackage{wrapfig}
\usepackage{color}
\usepackage{soul}
\usepackage{braket}
\usepackage{verbatim}
\usepackage{mathtools}
\usepackage{mathrsfs}
\usepackage{amsfonts}
\usepackage[english]{babel}
\usepackage{textcomp}
\usepackage{float}
\usepackage{comment}
\usepackage{url}

\usepackage{multirow}

\excludecomment{toexclude}
\usepackage{changepage} 

\begin{document}

\title{Topology and Martensitic Phase Transformations}
\author{M. Yin and D. D. Vvedensky}
\maketitle

\newpage
~\textbf{Abstract:} Triply periodic minimal surfaces (TPMS) are discovered to conform to surfaces of given charge density distributions embedded in crystals [Z. Kristallogr. \textbf{170}, 138 (1985)]. Based on our previous work [Phys. Rev. Mater. \textbf{9}, 073802 (2025)], we discovered that crystals can have surfaces of a given charge density converging to TPMS. We also discovered that end states connected by a martensitic phase transformation should have their corresponding TPMS being topologically equivalent. In this work, we gave an explanation for the topological continuity of a martensitic phase transformation and studied how TPMS indicate whether a non-magnetic crystal can undergo a martensitic phase transformation or not.\\

\textbf{keywords:} minimal surface, topology, martensitic phase transformations, symmetry

\section{Introduction}
\label{sec1}

\begin{figure}[t!]
	\centering
	\includegraphics[width=1\linewidth]{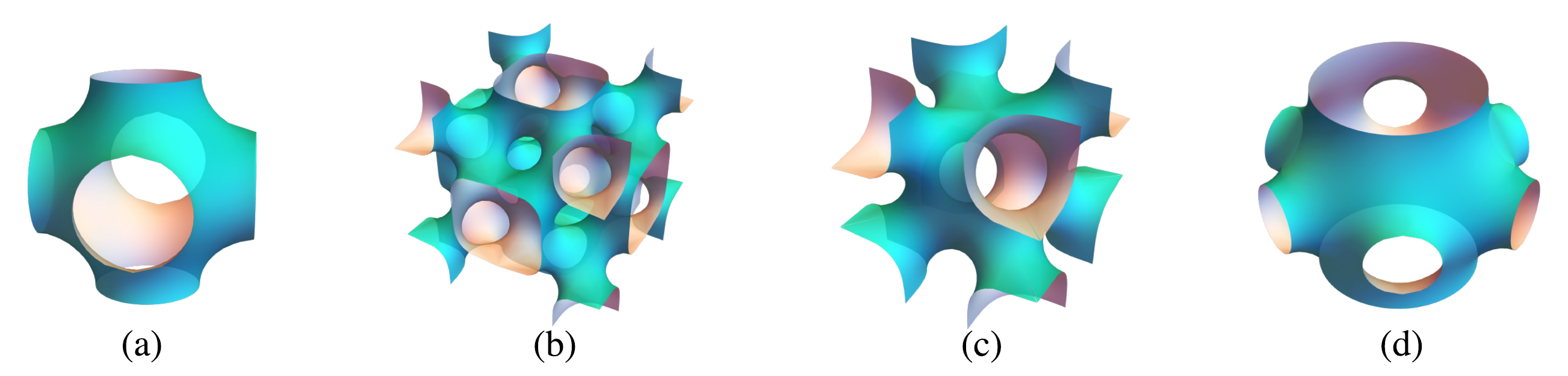}
	\caption{Examples of TPMS. (a) Schwarz P surface (P), (b) Schoen F-RD surface (F-RF), (c) Schoen I-WP surface (I-WP), (d) Schoen H$^\prime$-T surface (H$^\prime$-T). }
	\label{fig1}
\end{figure}

\begin{figure}[b!]
	\centering
	\includegraphics[width=1\linewidth]{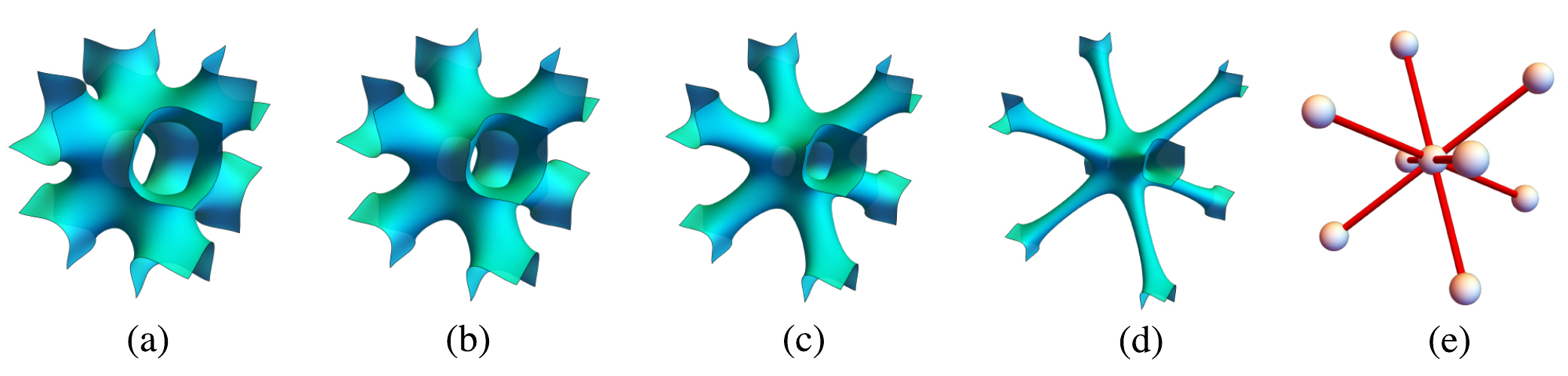}
	\caption{Example of forming a skeletal graph. (a) I-WP surface. From (b) to (d) shows a continuous shrinking of the I-WP surface. (e) Ball-stick illustration of the skeletal graph of the I-WP surface. Spheres represent ends and sticks represent arms connecting two ends.}
	\label{fig0}
\end{figure}

\begin{figure}[t!]
	\centering
	\includegraphics[width=1\linewidth]{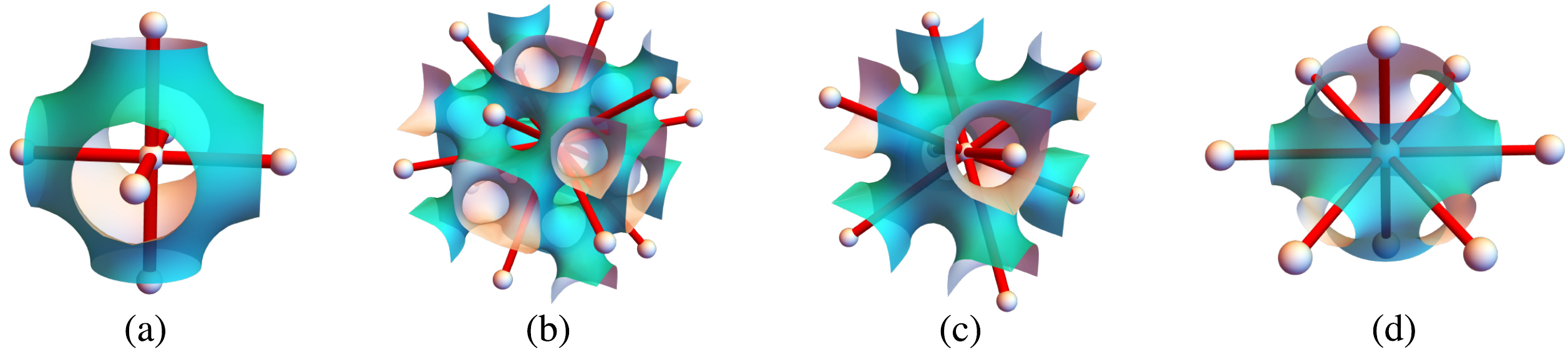}
	\caption{From (a) to (d) are skeletal graphs of the P, F-RD, I-WP and H$^\prime$-T surfaces, respectively. Balls represent ends and sticks represent arms connecting two ends.}
	\label{fig3}
\end{figure}

Minimal surfaces are defined as smooth surfaces with their mean curvature $M=-(k_1+k_2)/2$ vanishing everywhere, where $k_1,k_2$ are principal curvatures \cite{carmo16}. Catenoids and helicoids are two examples of minimal surfaces. If a smooth surface has its Gaussian curvature $K=k_1k_2$ zero everywhere, then this surface is said to be flat, such as a plane. A minimal surface with triple periodicity, such that it can be formed by parallel-transporting its surface unit along three mutually independent directions, is called a triply periodic minimal surface (TPMS) \cite{hyde96} (Fig.~\ref{fig1}). One interesting feature about TPMS is that they resemble surfaces of a given charge density in crystals, whether exchange-correlations are included or not \cite{schnering87,schnering91,nesper85,yin25}. The point group of crystals is a subgroup, either proper or improper, of the point group of their corresponding TPMS. Since the point group of the crystal is a subgroup of its corresponding Bravais lattice, the TPMS is analogous to the Bravais lattice, which gives the basic structure and periodicity of a crystal. From the perspective of topology, TPMS having the same genus, which can be understood as the number of holes on a surface, can deform into each other continuously, \textit{i.e.}, without tearing, piercing or penetrating the surface \cite{fomenko00}.~By continuously deforming, \textit{i.e.}, shrinking and expanding a TPMS, one can obtain a connected graph named as the skeletal graph of this TPMS \cite{schoen70,gupta18}. Such a skeletal graph is formed by edges and vertices, as shown in Fig.~\ref{fig0} \cite{schoen70,gupta18}. Skeletal graphs corresponding to TPMS shown in Fig.~\ref{fig1} are given in Fig.~\ref{fig3}.
Martensitic phase transformations are displacive solid-to-solid phase transformations during which atomic displacements are smaller than the lattice spacing \cite{battacharya03,christian02,olson81}. Iron \cite{pereloma12}, NiTi \cite{buehler63}, and Zr \cite{hao08} are examples of crystals proven experimentally to exhibit martensitic phase transformations. Martensitic phase transformations have been studied in various ways by researchers, including Landau's theory of phase transformations \cite{toledano96,izyumov94,falks1980}, analysis based on molecular dynamics and density functional theory (DFT) calculations \cite{lagoudas08,karewar18}, and modelling of shears  \cite{bain24,olson72,nishiyama78,sachs30} such as the Bain deformation. Martensites that can restore their original shape when heated up are named shape-memory materials, such as NiTi. Studies of energy barriers occurring during martensitic phase transformations discovered that there exist barrierless paths for the shape memory alloys but not for ordinary martensites \cite{hatcher09b,hatcher09,vishnu10,kibey09,wang21,morris06,krcmar20,niu16,okatov09,wang21,zhang21}.

In this work, we consider the topological continuity of martensitic phase transformations discovered in our previous work \cite{yin25} using skeletal graphs. We used examples of the BCC--FCC phase transformation modelled using the Bain deformation, the Nishiyama--Wassermann (NW) deformation and the Bogers--Burgers (BB) deformation, as well as the BCC--HCP (hexagonal) and BCC--monoclinic martensitic phase transformations to clarify our statements.  
\section{Bravais Lattice and TPMS}
\label{sec2b}

According to our previous work \cite{yin25}, we have found that the surface of a given charge density should converge to TPMS, such that the point group of this crystal is a subgroup, either proper or improper, of the point group of this TPMS. This is similar to the relationship between the point groups of a crystal and its corresponding Bravais lattice. The Bravais lattice of a crystal consists of lattice points. A lattice point is a union of atoms, and all these unions have the same chemical and symmetrical environment. Notice that we are talking about non-magnetic materials here since magnetism will lead to variation in the crystal symmetry. For example, without magnetism, the $\alpha$-Fe is in space group $Im\overline{3}m$, while including the magnetism, since the $\alpha$-Fe is ferromagnetic, the space group will be reduced to $I4/mmm$.  
\begin{figure}[b!]
	\centering
	\includegraphics[width=1\linewidth]{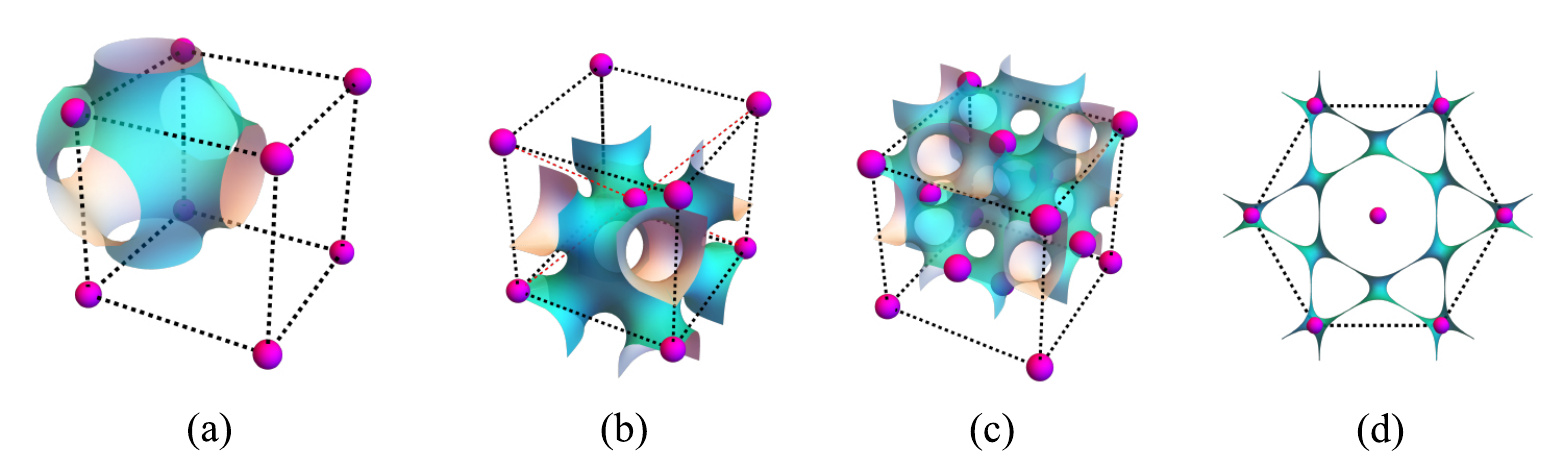}
	\caption{Sketch of how lattice points locate at flat points on a TPMS. (a) NiTi lattice (simple cubic) on the P surface, (b) BCC lattice on the I-WP surface, (c) FCC lattice on the F-RD surface, (d) hexagonal lattice on the H$'$-T surface. }
	\label{fig15}
\end{figure}
As shown in Fig.~\ref{fig15}, we can locate the Bravais lattice of a crystal on its corresponding TPMS using points having the same (zero) Gaussian curvature such that the lattice structure (lattice angles and ratios of lattice parameters) and lattice periodicity are preserved. Magenta balls in Fig.~\ref{fig15} are lattice points of the corresponding Bravais lattice. Flat points are chosen here to be lattice points because they can reveal both the lattice periodicity and the structure symmetry. Take B2 NiTi as an example. As shown in Fig.~\ref{fig18}(a,c), if we parallel transport the P surface in Fig.~\ref{fig15}(a) to the centre of B2 NiTi and then connect Ni atoms to Ti atoms, we see that the lines penetrate the P surface at a flat point. Repeating connecting atoms at the corner to the one at the centre, we see a structure formed by these crossing points on the P surface, as shown in Fig.~\ref{fig18}(b,e). Points in Fig.~\ref{fig18}(b) are related to each other by symmetry operations, \textit{e.g.}, a 4-fold rotational symmetry, belonging to the $O_h$ group. Any one of these 8 points can be chosen to be the start point to construct a simple cubic Bravais lattice [Fig.~\ref{fig18}(c,e)].
\begin{figure}[t!]
	\centering
	\includegraphics[width=.9\linewidth]{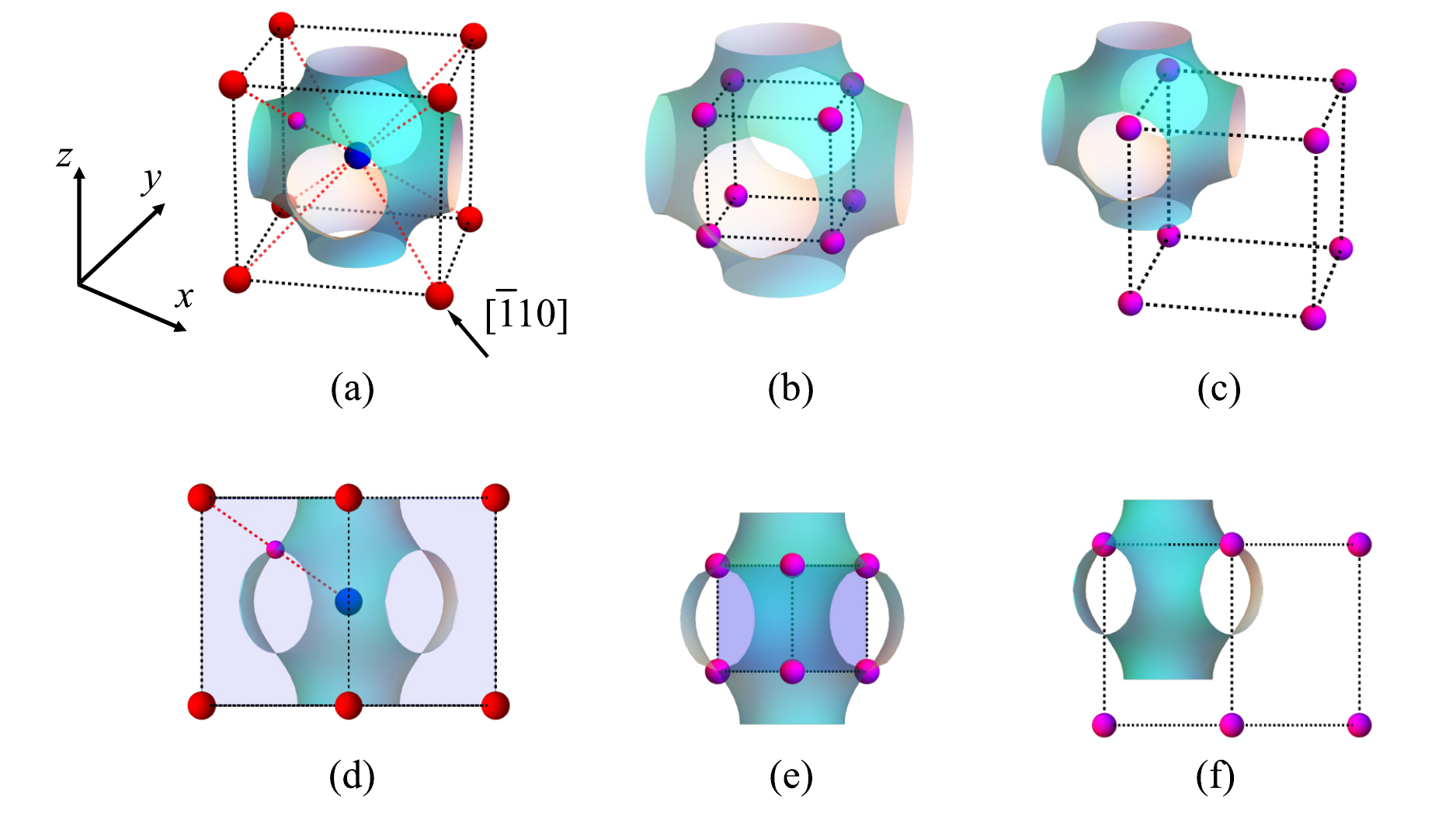}
	\caption{(a) Sketch of a flat point on P crossing which the line connecting a Ni atom (red sphere) and a Ti atom (blue sphere) penetrating the P surface. (b) Sketch of all flat points obtained by connecting one atom at the conner to the one at the centre. (c) Sketch of the Bravais lattice of B2 NiTi, with the start point chosen to be the one given in (a). (d,e,f) are the the view of (a,b,c) form the $[\overline{1}10]$ direction, respectively. }
	\label{fig18}
\end{figure}
\noindent
Moreover, if we look at the connection between lattice points and the corresponding skeletal graph of this crystal, as shown in Fig.~\ref{fig10}, we see that the connection resembles the skeletal graph (Fig.~\ref{fig3}). Hence, we can see that a crystal (non-magnetic) should have its corresponding TPMS satisfying:
\begin{enumerate}
	\item There exists a group-subgroup relation between the point groups of the TPMS and the crystal;
	\item The Bravais lattice of the crystal can be located at points of the same Gaussian curvature, such as flat points, on this TPMS, as long as the lattice structure and periodicity are preserved;
	\item The connection between lattice points of a crystal should resemble the skeletal graph of its corresponding TPMS.
\end{enumerate}
\noindent
Based on the DFT calculations given in \cite{yin25}, Table~\ref{table1} lists
symmetry of materials and their corresponding TPMS. According to Table~\ref{table1}, we can see that a martensitic phase transformation has its end phases described by TPMS having the same genus, while an ordinary phase transformation does not. To be specific, the FFC--BCC phase transformation in Al/Cu is not martensitic, and it witnesses a decrease in the genus from 6 to 4. Phase transformations such as HCP--hexagonal in Zr, HCP--BCC in Zr and BCC--HCP in Na are martensitic \cite{young75,straub71,hao08}, in which the genus is conserved.
\begin{figure}[b!]
	\centering
	\includegraphics[width=.9\linewidth]{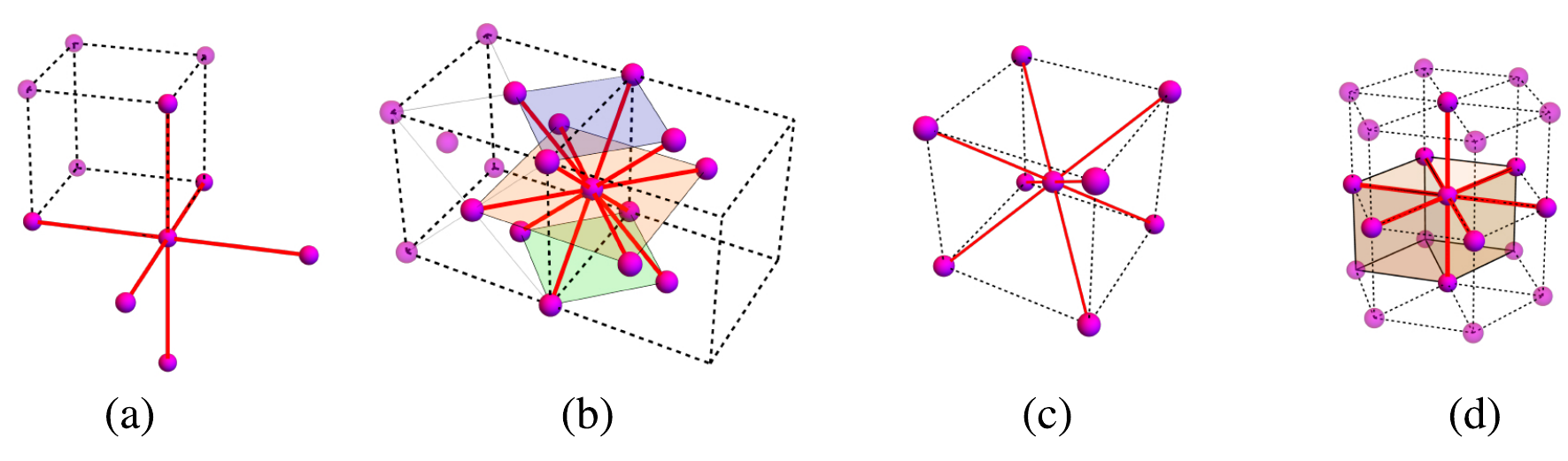}
	\caption{Sketch of connections between one lattice point and its nearest neighbours in (a) a simple cubic lattice, (b) an FCC lattice, (c) a BCC lattice and (d) an HCP lattice. Magenta spheres are for lattice points and red lines are for the arms. }
	\label{fig10}
\end{figure}
\begin{table}[b!]
	\centering
	\caption{List of symmetry groups of different materials and their corresponding TPMS. ``MT'' stands for ``martensitic phase transformation'' and ``SMA'' stands for ``shape memory alloy''.}
	\label{table1}
\resizebox{\textwidth}{!}{	\begin{tabular}{l @{\quad} c  @{\quad} c  @{\quad} c @{\quad}  l @{\quad} c @{\quad}  c @{\quad} c  @{\quad}  c  @{\quad} c @{\quad} c @{\quad} c @{\quad} c @{\quad} c@{\quad} c @{\quad} c }
		\toprule
	\multirow{2}{*}{\textrm{Material}} & \multicolumn{5}{c}{\textrm{Initial}}&  & \multicolumn{5}{c}{\textrm{Final}}& &\multirow{2}{*}{\textrm{MT}} & \multirow{2}{*}{\textrm{SMA}} \\
	\cline{2-12}
	& \textrm{SG }&\textrm{PG } & \textrm{TPMS} & genus & \textrm{PG }  & &\textrm{SG} & \textrm{PG } & \textrm{TPMS} & genus & \textrm{PG } &\\
	\hline
		Al \cite{mishin99} & $Fm\overline{3}m$ &$O_h$& F-RD &6  &$O_h$ &&  $Im\overline{3}m$ &$O_h$ & I-WP & 4  &$O_h$ && $\times$ & $\times$\\
		Cu  \cite{mishin01} & $Fm\overline{3}m$ & $O_h$& F-RD & 6 &$O_h$ & &$Im\overline{3}m$ &$O_h$ & I-WP & 4  &$O_h$ && $\times$ & $\times$\\
		K \cite{degtyareva14} & $Im\overline{3}m$ & $O_h$ & I-WP & 4& $O_h$  &&$Fm\overline{3}m$ & $O_h$ & F-RD & 6& $O_h$ & &   $\times$ & $\times$\\
		ClCs \cite{watanabe77} & $Pm\overline{3}m$ &$O_h$ & P & 3 & $O_h$ & &$Fm\overline{3}m$ & $O_h$ & F-RD & 6 & $O_h$ & &   $\times$ & $\times$\\
		Pb \cite{kuznetsova02} &   $Fm\overline{3}m$ &$O_h$& F-RD &6  &$O_h$ &&  $P6_3/mmc$ &$D_{6h}$ & H$'$-T & 4  &$D_{6h}$ && $\times$ & $\times$\\
		\hline
		Na \cite{suzuki83} & $Im\overline{3}m$ &$O_h$ &I-WP & 4 & $O_h$ & & $P6_3/mmc$ &$D_{6h}$& H$^\prime$-T &6  &$D_{6h}$  && \checkmark & $\times$\\
		\multirow{2}{*}{Zr \cite{ghosh14,hao08}} &\multirow{2}{*}{ $P6_3/mmc$} & \multirow{2}{*}{$D_{6h}$ } & \multirow{2}{*}{ H$^\prime$-T }& \multirow{2}{*}{4} & \multirow{2}{*}{ $D_{6h}$ } &&  $P6/mmm$ & $D_{6h}$ & H$^\prime$-T & 4  & $D_{6h}$ &&\checkmark &$\times$ \cr
		&&&&&&& $Im\overline{3}m$ &$O_h$& I-WP & 4 &$O_h$&&\checkmark &$\times$\\
		Ti \cite{trinkle03,hennig05} &  $P6_3/mmc$  & $D_{6h}$ & H$^\prime$-T & 4  & $D_{6h}$ & & $P6/mmm$ & $D_{6h}$ & H$^\prime$-T & 4  & $D_{6h}$ &&\checkmark &$\times$\cr 
		Li \cite{bollinger11} &  $Im\overline{3}m$ &$O_h$ &I-WP & 4 & $O_h$ & & $P6_3/mmc$ &$D_{6h}$& H$^\prime$-T &6  &$D_{6h}$  && \checkmark & $\times$\\
		\hline
		\multirow{5}{*}{Ni-Ti  \cite{hatcher09b,hatcher09,liu20}} & \multirow{5}{*}{$Pm\overline{3}m$} & \multirow{5}{*}{$O_h$} & \multirow{5}{*}{P} & \multirow{5}{*}{3 } &\multirow{5}{*}{$O_h$}&&$P2_1/m$ & $C_{2h}$ & oPb-family&3  & $C_{2h}$ &&\checkmark &\checkmark\cr
		&&&&&&& $P4/mmm$ &$D_{4h}$ & $tP$-family & 3 &$D_{4h}$ &&\checkmark &\checkmark\cr
		&&&&&&& $Pcmm$ &$D_{2h}$ & oPa & 3& $D_{2h}$ & &\checkmark &\checkmark \cr
		&&&&&&& $Cmcm$ &$D_{2h}$ & oPa & 3& $D_{2h}$  & &\checkmark &\checkmark \cr
		&&&&&&&  $P3,\,P\overline{3}$ & $C_3,\, C_{3i}$ &oPb(H)& 3 &$D_{3d}$ &&\checkmark &\checkmark \\
		\multirow{3}{*}{Au-Cd \cite{gandi23,lee17}} & \multirow{3 }{*}{$Pm\overline{3}m$} & \multirow{3 }{*}{$O_h$} & \multirow{3 }{*}{P} & \multirow{3 }{*}{3}&\multirow{3 }{*}{ $O_h$} & & $P3$ & $C_3$ & oPb(H) & 3 & $D_{3d}$ && \checkmark  &\checkmark \cr
		&&&&&&& $Pmma$ & $D_{2h}$ &oPa & 3 & $D_{2h}$ & & \checkmark &\checkmark\cr
		&&&&&&& $Cmcm$ & $D_{2h}$ & oPa & 3 & $D_{2h}$ && \checkmark  &\checkmark \\ 
		  Ag-Cd \cite{masson58} & $Pm\overline{3}m$ & $O_h$ & P & 3& $O_h$ &  & $Cmcm$ & $D_{2h}$ & oPa & 3 & $D_{2h}$  && \checkmark  &\checkmark  \\
		Ti-Pd \cite{lee17} &  $Pm\overline{3}m$ & $O_h$ & P & 3 & $O_h$  & & $P2_1/m$ & $C_{2h}$ & oPb-family & 3 & $C_{2h}$  & & \checkmark & \checkmark\\
		  Ti-Au \cite{lee17} &  $Pm\overline{3}m$ & $O_h$ & P & 3 & $O_h$ & &  $Pmma$ & $D_{2h}$ &oPa & 3 & $D_{2h}$ & & \checkmark &\checkmark\\
		  Ti-Pt \cite{lee17} &  $Pm\overline{3}m$ & $O_h$ & P & 3 & $O_h$  & & $P2_1/m$ & $C_{2h}$ & oPb-family & 3 & $C_{2h}$  & & \checkmark & \checkmark\\
		  Ni-Al \cite{maznoy19} & $Pm\overline{3}m$ & $O_h$ & P & 3 & $O_h$ & &$P4/mmm$ &$D_{4h}$ & $tP$-family & 3 &$D_{4h}$ &&\checkmark &\checkmark \\
		  Cu-Sn \cite{leineweber23} & $Pm\overline{3}m$ & $O_h$ & P & 3  & $O_h$ & & $Pmmn$ & $D_{2h}$ &oPa & 3 & $D_{2h}$ & & \checkmark &\checkmark \\
		\hline
\end{tabular}}
\end{table}

\section{Conservation of Genus and Explanation}
\label{sec2}

\begin{figure}[t!]
	\centering
	\includegraphics[width=.6\linewidth]{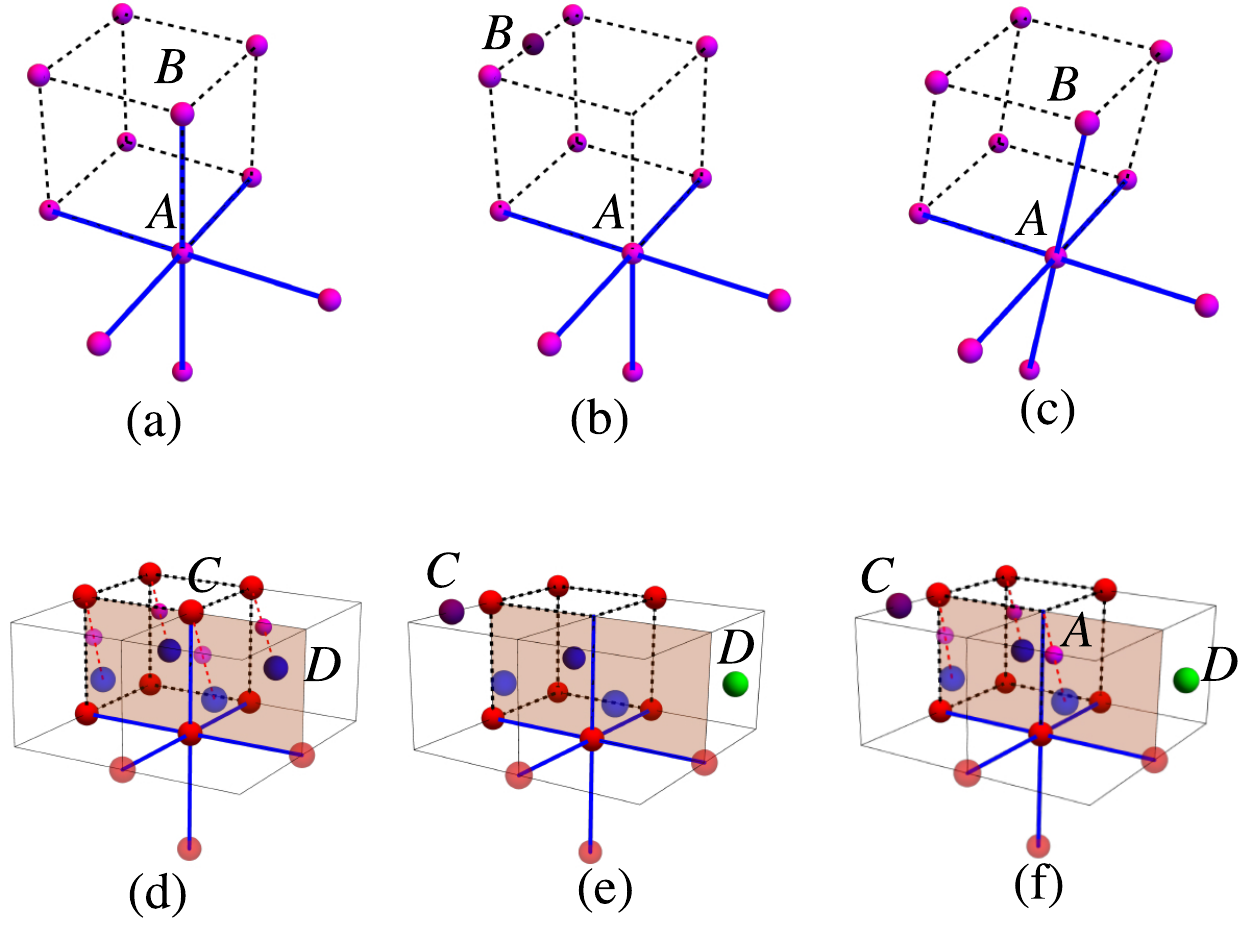}
	\caption{Sketch of variation in connections between lattice points in (a,b,c) a monatomic crystal and (d,e,f) a diatomic crystal. Magenta spheres are for lattice points, red and blue spheres are for atoms and blue lines are for connections between one lattice point and its nearest neighbours. Green and purple spheres are for displaced red and blue atoms, respectively.}
	\label{fig9}
\end{figure}

Such a conservation of or variation in genus can be understood from the perspective of the lack of diffusion in martensitic phase transformations. If the phase transformation is diffusionless, then we can expect no change in the connection between one lattice point and its neighbours. The ``neighbours'' means the nearest lattice points along each basis vector. For instance, in a simple cubic system, the neighbours connected to the lattice point located at the origin are at $\pm\bm{a}_i,\, i=1,2,3$, $\bm{a}_i$ are basis vectors of the simple cubic lattice. Beginning with a monatomic crystal, each of whose lattice points consists of one atom. As shown in Fig.~\ref{fig9}(a,b), if the phase transformation is not diffusionless, then the diffusion will lead to a breaking of the connection between the central lattice point $A$ and its neighbour $B$, given $B$ is not one of the nearest neighbours of $A$. For a diatomic system given in Fig.~\ref{fig9}(c), for instance, the diffusion will lead to a displacement of atoms $C$ and $D$ [Fig.~\ref{fig9}(d)]. Hence, in the new lattice [Fig.~\ref{fig9}(d)], connections between lattice point $A$ and its nearest neighbours will change. Diffusionless progress is realised by a shear, which is a relative displacement between two adjacent atomic layers. As shown in Fig.~\ref{fig9}(c), the connection between lattice point $A$ and its neighbours is preserved.

\begin{figure}[t!]
	\centering
	\includegraphics[width=1\linewidth]{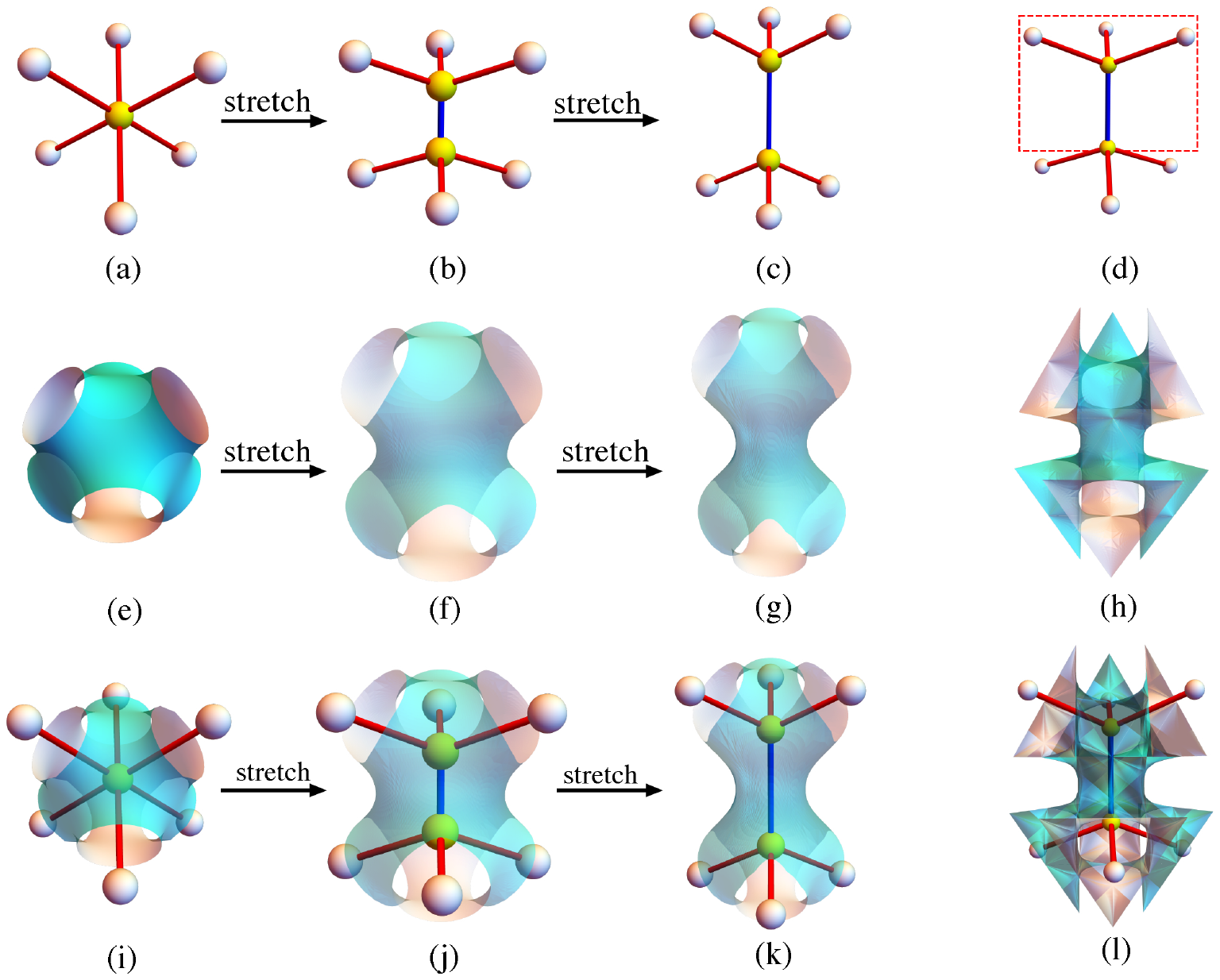}
	\caption{Sketch of continuously deforming (a) the skeletal graph of the P surface to (d) the skeletal graph of the D surface. (e,f,g,h) are the surface deformations corresponding to (a), (b), (c) and (d). (i), (j), (k) and (l) are combinations of surfaces and their skeletons during the deformation. The skeletal graph of (g) is the same as the one of (h), the D surface and by a further continuous deformation, one can obtain (h) from (g).}
	\label{fig11}
\end{figure}

 Since the connection between lattice points resembles the corresponding skeletal graph of this crystal, an invariance of arms connecting one lattice point to its nearest neighbours means an invariance in the skeletal graph corresponding to the crystal. Recalling that the skeletal graph of a crystal is obtained by continuously deforming a TPMS describing this crystal, we see such an invariance means a conservation of genus. Equivalently, surfaces that are topologically equivalent also have topologically equivalent skeletal graphs. For instance, the P surface can continuously deform into the D surface. By looking at their skeletal graph, we see that by stretching one end and then twisting the skeletal graph of the P surface, we will obtain the skeletal graph of the D surface, as shown in Fig.~\ref{fig11}. It is important to note that "stretching" a lattice point does not imply splitting it in real space, but rather, it refers to the reunion of lattice points from the previous phase with those from the current phase. Furthermore, such a stretching will happen to all lattice points, not limited to some particular ones. Now, we would like to employ some examples to clarify our statements.

\section{Results}
\label{sec3}

Al and Cu can exhibit an FCC$-$BCC phase transformation \cite{fiquet19,polsin17,schmidt95}; Na is proven theoretically to undergo a BCC$-$HCP martensitic phase transformation when cooled under 36K \cite{young75,straub71}. Experimentally, martensite phases of Zr, whose austenite phase is in an HCP lattice, can either transform to a BCC or a hexagonal lattice \cite{hao08}. NiTi, as one of the most well-known shape-memory materials, is proven to exhibit a B2$-$B19$^\prime$ martensitic phase transformation experimentally \cite{hatcher09b,hatcher09}. In this section, we will analyse the BCC$-$FCC phase transformation in Cu/Al with their corresponding genus changing from 4 to 6 and BCC$-$HCP/hexagonal phase transformation in Zr/Na with their genus conserved using the method stated above. Then, we will gain more insight into the B2--B19$'$ phase transformation in NiTi.

\subsection{BCC$-$FCC Transformation in Cu/Al}

The FCC--BCC phase transformation in Cu/Al involves a surface deformation from the F-RD surface (FCC) to the I-WP surface (BCC). As shown in Fig.~\ref{fig3}, their skeletal graphs have 12 (FCC) and 8 (BCC) ends, respectively. Hence, these two skeletal graphs cannot deform into each other continuously, and indeed, we see the F-RD surface having genus 6 while the I-WP surface has genus 4. Hence, we see that this phase transformation is not martensitic from the perspective of topology.

\subsection{Phase Transformations in Zr and Na}
 
\begin{figure}[b!]
	\centering
	\includegraphics[width=1\linewidth]{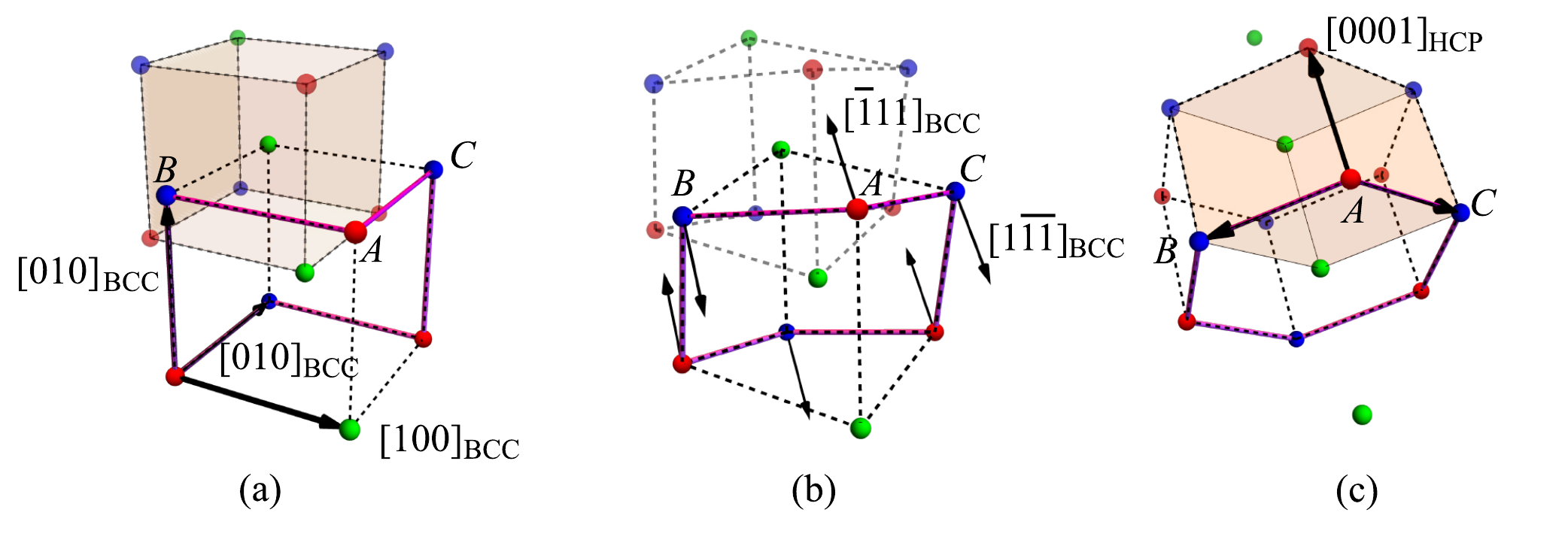}
	\caption{Sketch of the transformation from a BCC lattice to a hexagonal lattice. Atoms labelled red go along $[\overline{1}11]_{\text{BCC}}$, blue go along $[1\overline{1}\overline{1}]_{\text{BCC}}$ and green stay at their original positions. Shaded area in (a) shows unit cell and in (c) shows primitive cell.}
	\label{fig4}
\end{figure}                                                   

The BCC--HCP lattice transformation can be realised by the $[1\overline{1}\overline{1}]_{\text{BCC}}$ and $[\overline{1}11]_{\text{BCC}}$ shear with a distance $a/6$, where $a$ is the lattice parameter of the original BCC lattice (Fig.~\ref{fig4}), such that we can see $[\overline{1} 11]_{\text{BCC}}\parallel [0001]_{\text{hex}}$. With this shear, we have the skeletal graph corresponding to the I-WP surface deforming continuously to the skeletal graph of the H$'$-T surface. Moreover, topologically, the I-WP and the H$'$-T surfaces all have genus 4. Thus, topologically, the BCC$-$hexagonal transformation in Zr can be martensitic. Since the HCP lattice has the same Bravais lattice as the hexagonal lattice, the HCP--hexagonal martensitic phase transformation is also permitted topologically. 

\subsection{BCC$-$Monoclinic Transformation in NiTi}
\begin{figure}[t!]
	\centering
	\includegraphics[width=1\linewidth]{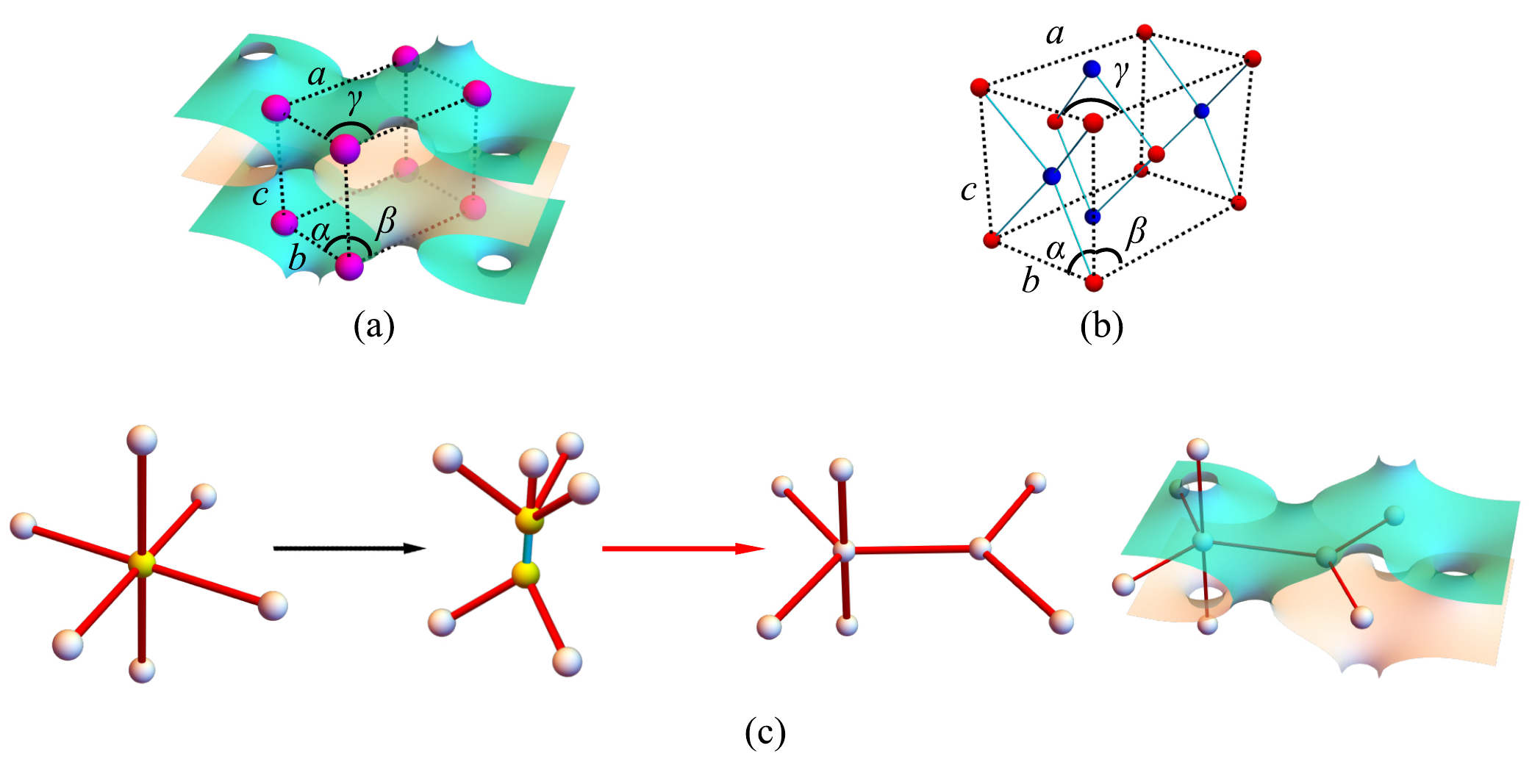}
	\caption{Sketch of (a) Bravais lattice of the B19$'$ NiTi against the corresponding oPb TPMS, (b) crystal lattice of the B19$'$ NiTi and (c) a continuous deformation from the skeletal graph of the B2 NiTi to the B19$'$ NiTi.}
	\label{fig16}
\end{figure}

\begin{figure}[t!]
	\centering
	\includegraphics[width=1\linewidth]{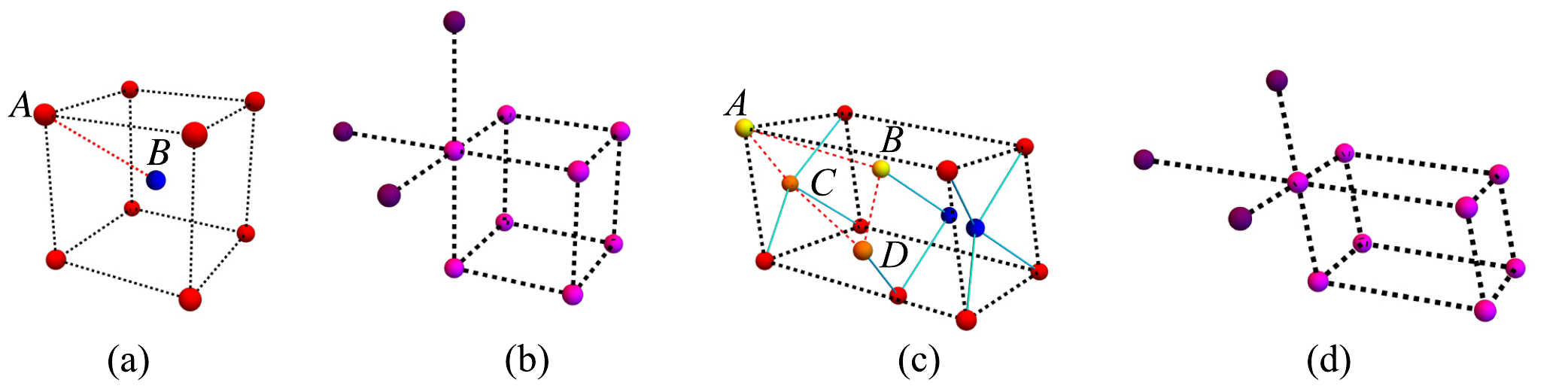}
	\caption{Sketch of (a) crystal lattice of the B2 NiTi, atoms $A$ (Ni) and $B$ (Ti) form a lattice point, (b) Bravais of the B2 NiTi, (c) crystal lattice of the B19$'$ NiTi, atoms $A,B$ (Ni) and $C,D$ (Ti) form a lattice point and (d) Bravais lattice of the B19$'$ NiTi. Red and blue spheres represent Ni and Ti atoms, respectively. Magenta and purple spheres represent lattice points. }
	\label{fig7}
\end{figure}

Under a martensitic phase transformation, the B2 ($Pm\overline{3}m$) NiTi can transform to the B19$'$ ($P2_1/m$) NiTi. According to Sec.~\ref{sec2b}, we found that the B19$'$ NiTi can be described by a surface belonging to the oPb-family \cite{chen20,chen21b} that can deform into the P surface continuously [Fig.~\ref{fig16}(a,b)]. Angles $\alpha,\beta,\gamma$ in Fig.~\ref{fig16}(a) equal to angles $\alpha,\beta,\gamma$ in Fig.~\ref{fig16}(b) and ratios of $a/b,a/c$ in Fig.~\ref{fig16}(a) equal to ratios of $a/b,a/c$ in Fig.~\ref{fig16}(b). Hence the lattice structure is preserved. The skeletal graph of such a surface can be obtained by stretching lattice points, as shown in Fig.~\ref{fig16}(c). Since the lattice point in the B19$'$ phase consists of 4 atoms, \textit{i.e.}, 2 Ni atoms and 2 Ti atoms, such a ``stretching'' indicates a union of 2 old lattice points chosen in the B2 phase (Fig.~\ref{fig7}). Therefore, the skeletal graph given in Fig.~\ref{fig16}(c) conforms to the connection between one lattice point and its nearest neighbours given in Fig.~\ref{fig7}(d). According to Fig.~\ref{fig7}(d), each lattice point in either the B2 NiTi or the B19$'$ NiTi has 6 nearest neighbours, which conforms to the topological equivalence of the skeletal graphs. Hence, the B2--B19$'$ martensitic phase transformation in NiTi is topologically permitted. 

\section{Conclusion}
\label{sec5}

In conclusion, if a crystal can undergo a martensitic phase transformation, then the TPMS corresponding to its end phases have the same genus, meaning they can continuously deform into one another. The topological continuity is attributed to the diffusionless nature of the phase transformation, which requires no breaking in connections between each lattice point and its nearest neighbours. As the connection between lattice points of a crystal is analogous to the skeletal graph that can be generated by continuously deforming the corresponding TPMS, topological continuity in connections implies topological continuity in the skeletal graph, and as a result, the TPMS. According to our DFT results \cite{yin25}, martensitic phase transformations in Zr and Na are topologically continuous, while non-martensitic phase transformations in Cu and Al are not. We also discovered that the Bravais lattice of a crystal can be positioned at points having the same Gaussian curvature, \textit{e.g.}, flat points, on its corresponding TPMS. Based on this, we found the B19$'$ NiTi can be located on a TPMS belonging to the oPb family that is topologically equivalent to the P surface. Each lattice point in the B19$'$ NiTi is connected to 6 nearest neighbours, consistent with the B2 NiTi. This conforms to the topological continuity of their TPMS (skeletal graphs), which is in accordance with the topological criteria of a martensitic phase transformation. Shape memory materials represent a specific type of martensites. Thus, as shown in Table.~\ref{table1}, shape memory materials can undergo a topologically continuous phase transformation. The differences in transformation paths between shape memory alloys and ordinary martensites have been discussed in Ref.~\cite{yin24}.

\bibliographystyle{unsrt}
\bibliography{paper.bib}

\end{document}